\documentclass[showpacs,preprintnumbers,amsmath,amssymb,floatfix]{revtex4}
\usepackage{graphicx}% Include figure files
\usepackage{dcolumn}% Align table columns on decimal point
\usepackage{bm}% bold math
\usepackage{psfig}
\begin{document}
\preprint{LU TP 02-09}
\preprint{NORDITA-2002-22 HE}

\title{Center--Vortex Solutions of the Yang--Mills\\Effective Action in Three and Four Dimensions}

\author{Dmitri Diakonov$^{1,2}$ and Martin Maul$^3$}
\affiliation{
{$^1$ NORDITA, Blegdamsvej 17, 2100 Copenhagen \O, Denmark}\\
{$^2$ St. Petersburg Nuclear Physics Institute, Gatchina,
St.Petersburg 188 300, Russia} \\
{$^3$ Department of Theoretical Physics, Lund University,
S\"olvegatan 14A, S - 223 62 Lund, Sweden}}

\date{\today}% It is always \today, today,

\begin{abstract}
We calculate the one-loop effective action of the $SU(2)$ Yang--Mills
theory for center-vortex configurations, both in $3d$ and $4d$. We find
that in both cases there are minima of the effective action,
corresponding to vortices of the transverse size approximately $4/g_3^2$
and $1.7/\Lambda_{\overline{MS}}$, respectively. The values of
the effective actions at the minima are {\it negative}, suggesting
that the Euclidian vacuum may be unstable with respect to creation of
vortices.
\end{abstract}

%\keywords{Suggested keywords}     %Use showkeys class option if keyword
\maketitle

\newcommand{\ur}[1]{(\ref{#1})}
\newcommand{\urs}[2]{(\ref{#1},\ref{#2})}
\newcommand{\urss}[3]{eqs.(\ref{#1},\ref{#2},\ref{#3})}
\renewcommand{\vec}[1]{{\bf #1}}
\newcommand{\eq}[1]{eq.~(\ref{#1})}
\newcommand{\eqs}[2]{eqs.~(\ref{#1},\ref{#2})}
\newcommand{\Eq}[1]{Eq.~(\ref{#1})}
\newcommand{\Eqs}[2]{Eqs.(\ref{#1}, \ref{#2})}
\newcommand{\eqss}[3]{eqs.~(\ref{#1},\ref{#2},\ref{#3})}
\newcommand{\e}{\epsilon}
\newcommand{\n}{\bf{n}}
\newcommand{\x}{\bf{x}}
\newcommand{\ee}[1]{e_{\{#1\}}}
\newcommand{\half}{\frac{1}{2}}

%****  Macros for Roman font in formulas

\def\Sp{\mbox{Sp}}
\def\Tr{\mbox{Tr}}
\def\det{\mbox{det}}
\def\Det{\mbox{Det}}

\section{Introduction}

Recently there have been certain indications from lattice simulations
that center-of-group vortices  \cite{'tHooft:1977hy,Mack}
may be responsible for the area behavior
of large Wilson loops, i.e. for confinement, see e.~g.~
\cite{Faber:2000um,DelDebbio:1996mh,ER} and references
therein. If long closed vortices populate the Euclidean vacuum of QCD,
resembling curved and entangled spaghetti in Italian {\em pasta}, it might
serve as a microscopic mechanism of confinement. \\

From the continuum side, however, vortices invoke more questions than there
are answers today. First of all, there are no finite-size vortex-type
solutions of the classical Yang--Mills equations in a non-compact space,
as the classical equations are dimensionless whereas the functional
to be minimized is the transverse energy with the dimension $1/{\rm cm}^2$.
Therefore, if one finds a solution, its dilatation to larger
size will have less energy. This shows that vortex-type solutions
of classical equations do not exist \cite{footnote}.
In a non-compact space the length scale and hence the vortex radius can be 
only set by quantum fluctuations. Therefore, vortices can at best
arise as saddle points of the effective action, with quantum fluctuations
about a trial vortex configuration taken into account. Second, if the
effective action contains vortex configurations as saddle points, their
energy per unit length in $3d$ or per unit surface in $4d$ should be
either small positive or, even better, slightly negative. If it is large
and positive there is no reason for the vortices to appear in the physical
vacuum, in the first place. Third, the typical lengths where vortices
bend to an angle of the order of unity must be on the average larger than
the characteristic transverse size of the vortex. If it is not so, the
fluctuation can be hardly called a vortex. Fourth, different vortices
in the vacuum or different segments of the same vortex need to be on
the average repulsive so that they do not glue up. If vortices are
permanently merging and splitting it is difficult to consider them as 
adequate degrees of freedom. A pilot study of the `spaghetti vacuum' in $4d$
has been performed many years ago by the Copenhagen group \cite{NNO,AO} who
considered a $2d$ lattice of parallel flux tubes and then presented arguments
that the lattice should ``melt'' into a disordered phase.
Ultimately, one has to build a consistent statistical mechanics of the
vortex vacuum in order to see that the `entangles spaghetti' have a preferred
vacuum free energy as compared to, say, the instanton liquid.
The statistical mechanics of instantons is quite a difficult problem
by itself \cite{Diakonov:1983hh,DM}, but vortices are far more difficult
to deal with.\\

In this paper, we address the first and in fact the easiest
questions from the above list. For vortices to be physical objects
and not artifacts of a regularization and gauge fixing, their
typical transverse sizes must be finite in physical units, i.e. to be
of the order of $1/g^2_3$ in $3d$ and of the order of
$1/\Lambda_{\rm QCD}$ in $4d$.  Correspondingly, the energy of a
vortex per unit length should be of the order of $g_3^2$ in $3d$,
and the energy per unit surface should be of the order of
$\Lambda_{\rm QCD}^2$ in $4d$. The appearance of $\Lambda_{\rm QCD}$
implies the transmutation of dimensions through renormalization; it
can be achieved only with quantum fluctuations about the vortex taken
into due account. We study here the effective action with quantum
fluctuations about a trial vortex configuration being integrated out,
at the one-loop level. Although the one-loop approximation is not the
whole story, it is the first attempt to find out if there are any
hints from the continuum Yang--Mills theory that isolated `thick'
vortices have the right to exist.
We seem to get a positive answer to this question. It seems
that quantized-flux vortices are dynamically preferred compared
with the perturbative vacuum.\\

%From the comparison of the energy of a center flux-1
%vortex with that of flux-zero vortex, it seems that quantized-flux
%vortices are dynamically preferred.\\

In what follows, we consider an idealized isolated infinitely long and
straight vortex perpendicular to the $(xy)$ plane. We choose an Ansatz
for the vortex profile in the transverse plane $\bar A_\mu(x,y)$
and expand the YM action into the classical part,
$\int \!d^dx F_{\mu\nu}^2(\bar A)/4g_d^2$, and the part quadratic in the
fluctuations $a_\mu$ about the background $\bar A_\mu$. Integration over
$a_\mu$ (and over ghost fields stemming from background gauge fixing)
is non-trivial, as the mere counting of levels as compared to the
free case is not obvious for the quantized-flux vortices; the
difficulty is related to the Aharonov--Bohm effect. We perform  this
integration by extending the Jost scattering theory (which `counts' the levels
properly) to the two-dimensional case: it results in the quantum part of the
effective action being finite after renormalization. The full effective action
is the sum of the classical and quantum parts. In fact, we study several
functional forms of the profile and find that in each case there exists a
minimum of the action with respect to the transverse spread of the vortex. It
turns out to be about $(4-5)/g_3^2$ in $3d$ and about
$(1.6-1.8)/\Lambda_{\overline{MS}}$ in $4d$. The values of the effective
action at the minima are {\em negative} in all cases considered. It indicates
that the Yang--Mills vacuum might be unstable in regard to the spontaneous
production of vortices. \\

This work has certain overlap with the earlier study by the Copenhagen
group \cite{NNO,AO} who started from the observation by Savvidy
\cite{Sav} that already a constant chromomagnetic field lowers the
Yang--Mills vacuum energy. The Savvidy vacuum has, however, a negative
fluctuation mode leading to instability. The Copenhagen group
has included the amplitude of the negative mode (interpreted
as a Higgs field but actually part of the gauge field) into the minimization
procedure and indicated  that the $2d$ hexagonal lattice of tubes with
quantized flux leads to a further lowering of vacuum energy. However neither
the fluctuations in the unstable mode nor in the higher modes were included.
In contrast, we take into full account all quantum fluctuations about a {\em
single} trial vortex. The full one-loop quantum calculation performed here
shows that already isolated center vortices in SU(2) and also their embedding
in SU(3) are energetically preferred compared with the perturbative vacuum.
This result comes both in $4d$ and $3d$ (the latter case has not been
considered in Refs. \cite{NNO,AO}) giving thus certain support to the
disordered-vortices scenario of confinement in both cases. \\

\section{The vortex Ansatz}
\label{sec1}

By definition of a $Z(N)$ vortex, the Wilson loop in the fundamental
representation of the $SU(N)$ gauge group, winding around the vortex
in the transverse plane assumes values being nontrivial elements of
the group center, as the radius of the Wilson loop tends to $\infty$:

\begin{equation}
{\rm P}\exp i \oint A_\mu dx^\mu
\rightarrow  \left( \begin{array} {ccc}
     e^{2\pi i k /N} && \\
   & \dots &          \\
   && e^{2\pi i k /N}
 \end{array}\right)\in Z_N,
\label{W1}
\end{equation}
where $k=1,\dots,N-1$. In this paper, we consider the $SU(2)$
gauge group (although we also make calculations in $SU(3)$ for
additional checks). In $SU(2)$ the only nontrivial element of the center
is the minus unity $2\times 2$ matrix. Taking for example a circular
Wilson loop of radius $\rho$ in the $(xy)$ plane around a vortex
centered at $\rho= \sqrt{x^2+y^2}=0$ we see that only the azimuthal
component of the Yang--Mills field is involved in \eq{W1},
$A_\phi^a=\epsilon_{\alpha\beta}n_\alpha A_\beta^a$, where
$n_\alpha$ is a unit vector in the $(xy)$ plane. One can always
choose a gauge where $A_\phi$ is independent of the azimuthal angle
$\phi$.  Generally speaking, it implies that the radial component
$A_\rho$ is nonzero, however, we shall neglect this component as the
condition \ur{W1} is compatible with $A_\rho=0$. If $A_\rho$ is
nonzero it can be reconstructed from gauge invariance by replacing
$\partial_\rho\to \partial_\rho\delta^{ab} + f^{acb}A_\rho^c$. A
circular Wilson loop in the $J=\frac{1}{2}$ representation lying in
the transverse plane and surrounding the vortex center is
\cite{Diakonov:1999gg}:

\begin{equation}
W_{\frac{1}{2}}(\rho)=\frac{1}{2}
\Tr\;{\rm P}\exp i \oint\rho A_\phi^a t^a d\phi=
\cos[\pi\mu(\rho)],\qquad
\mu(\rho)=\rho\sqrt{A_\phi^a(\rho) A_\phi^a(\rho)},
\label{W1/2}
\end{equation}
For an arbitrary representation of $SU(2)$, labeled by spin $J$, one
has:

\begin{equation}
W_J(\rho)=\frac{1}{2J+1}\frac{\sin[(2J+1)\pi\mu(\rho)]}
{\sin[\pi\mu(\rho)]}.
\label{WJ}\end{equation}
If $\mu(\rho)$ tends to an odd integer at large $\rho$, the Wilson loop
$W_J(\rho)\to (-1)^{2J}$. For half-integer representations it is minus
unity, hence $\mu(\infty)=2n+1$ is the condition that one deals with a
$Z(2)$ vortex. We shall only consider the case when $\mu(\infty)=1$.
For simplicity we assume that only one color component of $A_\phi$
is nonzero:

\begin{equation}
\bar A_\phi^a(\rho)=\delta^{a3}\frac{\mu(\rho)}{\rho},
\qquad \mu(0)=0,\qquad \mu(\infty)=1,
\label{Ansatz}\end{equation}
where $\mu(\rho)$ will be called the profile of the vortex. \Eq{Ansatz}
with all the rest components of the YM field set to zero, is the
Ansatz for the center vortex field we are going to investigate.

\section{The classical action of the vortex}

We take an idealized straight-line (in $3d$) or straight-surface (in $4d$)
vortex whose dimension in the longitudinal direction is
$L_\parallel^{(d-2)}$ ($d=3,4$), while in transverse plane it has a trial
profile $\mu(\rho)$. The classical action of the vortex
is (see Appendix A):

\[
S_{{\rm class}}=\frac{1}{4g_d^2}\int\!d^dx(F^a_{\mu\nu})^2
=\frac{1}{2g_d^2}\int\!d^{d-2}x_\parallel\!
\int\!\!d^2x_\perp\,(B^a_\parallel)^2
\]
\begin{equation}
=L_\parallel^{(d-2)}\;
\frac{\pi}{g_d^2}\!\int_0^\infty\!\!d\rho\:\rho
\left[\frac{1}{\rho}\partial_\rho(A_\phi^a\rho)\right]^2
=L_\parallel^{(d-2)}
\frac{\pi}{g_d^2} \int_0^\infty d\rho \rho \left[ \frac{1}{\rho}
\frac{\partial}{\partial \rho} \mu(\rho) \right]^2
\equiv L_\parallel^{(d-2)} {\cal E}_{\rm class}\;.
\label{classaction}\end{equation}
We remind the reader that in $3d$ the coupling constant $g_3^2$
has the dimension of mass, and the theory is convergent. In $4d$
the (bare) coupling $g_4^2$ is dimensionless but gets an infinite
renormalization from quantum fluctuations. To the one-loop accuracy
one can parameterize:

\begin{equation}
\frac{8\pi^2}{g_4^2}=\frac{11}{3}N\,\ln\frac{M}{\Lambda_{{\rm QCD}}}\;,
\label{bareg}\end{equation}
where $M$ is the ultra-violet cutoff in a particular regularization
scheme. The quantum part of the effective action is logarithmically
divergent; the renormalizability of the theory ensures that the
dependence on the UV cutoff $M$ is canceled in the full effective
action.

\section{Dimensional analysis}

Let us make a quick estimate of the `best' transverse size of the vortex
$\rho_0$ from a simple dimensional analysis. We denote the action
per unit longitudinal dimension of the vortex by ${\cal E}$ such that
$S=L_\parallel^{(d-2)}{\cal E}$, and call ${\cal E}$ the transverse
energy of the vortex. If $\rho_0$ is the scale of the profile
function $\mu(\rho)$ [e.g.~$\mu(\rho)=\exp(-\rho_0/\rho)$], we
find on dimensional grounds that in $4d$

\begin{equation}
{\cal E}_{{\rm class}}=\frac{A}{\rho_0^2}
\,\frac{11\,N}{24\pi^2}\,\ln\frac{M}{\Lambda_{{\rm QCD}}},
\label{c4de}\end{equation}
where $A$ is the dimensionless functional of the trial profile,

\begin{equation}
A=\pi\rho_0^2\int_0^\infty\!\frac{d\rho}{\rho}
\left(\frac{d\mu}{d\rho}\right)^2.
\label{A}\end{equation}
The renormalizability ensures that the quantum part of the effective
action has exactly the same coefficient in front of the divergent
$\ln M$ piece, but with the opposite sign, so that $\ln M$ cancels in
the sum of the classical and quantum parts:

\begin{equation}
{\cal E}_{{\rm quant}}= -\frac{A}{\rho_0^2}
\,\frac{11\,N}{24\pi^2}\,\ln(M \rho_0 B),
\qquad {\cal E}_{{\rm eff}}={\cal E}_{{\rm class}}
+{\cal E}_{{\rm quant}}=-\frac{A}{\rho_0^2}
\,\frac{11\,N}{24\pi^2}\,\ln(\Lambda_{{\rm QCD}}\rho_0 B),
\label{e4de}\end{equation}
where $A,B$ are numerical coefficients depending on the
concrete functional form of the profile $\mu(\rho)$.
${\cal E}_{{\rm eff}}$ has the minimum at

\begin{equation}
\rho_0^{\rm min}=\frac{\sqrt{e}}{\Lambda B },\qquad
{\cal E}^{{\rm min}}_{{\rm eff}}
= -\Lambda^2\frac{11\,N}{24\pi^2}\,\frac{A B^2}{2e},\qquad e=2.71828...
\label{m4de}\end{equation}
In $3d$ a similar dimensional analysis yields:

\begin{equation}
{\cal E}_{{\rm class}}=\frac{A}{g_3^2\rho_0^2}, \qquad
{\cal E}_{{\rm quant}}=-\frac{C}{\rho_0},
\label{e3de}\end{equation}
with the same coefficient $A$ as in $4d$. The sign of $C$ is apriori
unknown. However, since the theory is super-renormalizable one
can suppose that the sign of the quantum energy is the same as in $4d$ at
$M\bar\rho\to\infty$, i.e. negative (direct calculations below
confirm the expectation). The sum ${\cal E}_{{\rm eff}}={\cal
E}_{{\rm class}}+{\cal E}_{{\rm quant}}$ has then a minimum at

\begin{equation}
\rho_0^{\rm min}=\frac{1}{g_3^2}\,\frac{2A}{C},\qquad
{\cal E}^{{\rm min}}_{{\rm eff}}=-g_3^2\,\frac{C^2}{4A}.
\label{m3de}\end{equation}
Notice that both in $3d$ and $4d$ we get a natural result for the
transverse size of the vortex and a {\em negative} sign for its
transverse energy at the minimum. \\

In a sense we have already proven the main statements of the paper
put in the Abstract and the Introduction: these statements follow
from dimensional analysis and the renormalizability of the theory,
{\em provided} the coefficients $A,B,C$ are finite, which is not
obvious beforehand. In the rest of the paper we compute these
coefficients for different cases, and look for the best profile function
$\mu(\rho)$.

\section{The quantum action about a vortex}
\subsection{Definition of the quantum action}

To get the quantum action for $\mu(\rho)$ we integrate
over quantum fluctuations about a given background field
$\bar A_\phi^a(\rho)$ (see \eq{Ansatz}) considered to be a slowly
varying field with momenta $k_\perp$ up to certain $k_{\rm min}$.
Accordingly, quantum  fluctuations have momenta $k_\perp$ above
$k_{\rm min}$ and arbitrary longitudinal momenta. Writing $A_\mu=\bar
A_\mu + a_\mu$ we expand $F^2_{\mu\nu}(A+a)$ in the fluctuation field
$a_\mu$ up to the second order appropriate for the 1-loop
calculation. The term linear in $a_\mu$ vanishes due to the
orthogonality of high and low momenta and also because $\bar A_\mu$
is in our case independent of $\phi$ and longitudinal coordinates
whereas $a_\mu$ is dependent on those coordinates. \\

The quadratic form for $a_\mu$ is the standard
(see e.g. \cite{Diakonov:1983dt})

\begin{equation}
W^{ab}_{\mu\nu}=-[D^2(\bar A)]^{ab}\delta_{\mu\nu}-2f^{acb}
F^c_{\mu\nu}(\bar A),
\label{quadrform}\end{equation}
if one imposes the background Lorentz gauge condition,
\begin{equation}
D_\mu^{ab}(\bar A) a_\mu^b = 0,\qquad
D_\mu^{ab}(\bar A) = \partial_\mu\delta^{ab}+f^{acb}\bar A_\mu^c.
\label{backgauge}\end{equation}
The quantum part of the effective action is, as usually, given by the
small-oscillation determinants:

\begin{equation}
S_{{\rm quant}}[\bar A]= S_{{\rm gluon}}+S_{{\rm ghost}}
= \frac{1}{2}\ln\det(W^{ab}_{\mu\nu}) -\ln\det(-D_\mu^2).
\label{effact}\end{equation}
Subtraction of free determinants (with zero background field) is
understood. \\

The full effective action is
$S_{{\rm eff}}=S_{{\rm class}}+S_{{\rm gluon}}+S_{{\rm ghost}}$.
Since all terms are proportional to the longitudinal dimensions
of the vortex $L_\parallel^{(d-2)}$ we shall divide all terms by this
quantity; it will then become an equation for the appropriate
contributions to the transverse energy of the vortex,

\begin{equation}
{\cal E}_{{\rm eff}}={\cal E}_{{\rm class}}+{\cal E}_{{\rm gluon}}
+{\cal E}_{{\rm ghost}},
\label{Eefffull}\end{equation}
where all terms are certain functionals of the trial vortex profile
$\mu(\rho)$.

\subsection{The ghost operator}

For the vortex Ansatz \ur{Ansatz} the ghost operator $(D^2)^{ab}$ takes
the following simple form in the cylindrical coordinates
\cite{Diakonov:1999gg}:

\begin{equation}
D^2 = \left(\frac{1}{\rho}\frac{\partial}{\partial \rho} \rho
\frac{\partial}{\partial\rho}+\partial_3^2+\dots\partial_d^2\right)
\delta^{ab}
+ \frac{1}{\rho^2} \left(\delta^{ac}\frac{\partial}{\partial\phi}
+ f^{a3c} \mu(\rho)\right)
\left(\delta^{cb}\frac{\partial}{\partial \phi}
+ f^{c3b} \mu(\rho)\right).
\label{DD}\end{equation}
The eigenvalues of the structure constants $if^{a3c}$ in the color
space are

\begin{equation}
\alpha_c =\left\{ \begin{array}{cc}
 \{0, 1,-1\}  &  {\rm for \;\; SU(2),}  \\
\\
 \{0,0,1,-1,1/2,1/2,-1/2,-1/2\}  & {\rm for \;\; SU(3).}
\end{array} \right.
\label{eigalpha}\end{equation}

It is natural to choose a plane wave basis
$\exp(ik_\parallel \cdot x_\parallel)$ for the longitudinal
directions $x_3,\dots x_d$, and a polar basis $\exp(im
\phi)Z^{(c)}_m(\rho)$ for the transverse directions. The
eigenfunctions $Z^{(c)}_m(\rho)$ satisfy the second-order
differential equations depending on the color polarization
$c=1,\dots,(N_c^2-1)$:

\begin{equation}
\left[-\frac{1}{\rho}\frac{\partial}{\partial\rho}
\rho \frac{\partial}{\partial\rho}+\frac{(m-\alpha_c\,\mu(\rho))^2}{\rho^2}
\right]Z^{(c)}_m(\rho)=k_{\perp\,cm}^2Z^{(c)}_m(\rho).
\label{eig1}\end{equation}
For each color polarization $c$ one has to solve a separate eigenvalue
equation with a corresponding coefficient $\alpha_c$ from
\eq{eigalpha}.  The spectrum in $k_\perp$ is continuous with the
spectral density depending on $m$ and $c$. By putting the system into
a large circular box of radius $R$, one makes the spectrum
discrete, and the eigenvalues $k_\perp$ can be labeled by a discrete
number $n$. \\

The eigenvalues of the full operator $-D^2$ are then

\begin{equation}
E^2(c,m,n,k_\parallel)=k_\parallel^2+k_{\perp\,cmn}^2,
\label{E1}\end{equation}
and one has to sum over the color polarizations labeled by $c$, the
magnetic quantum  numbers $m$, the radial quantum numbers $n$ and
integrate over the continuous spectrum in $k_\parallel$. In the free
case ($\mu=0$) the eigenfunctions of \eq{eig1} are the ordinary
Bessel functions $J_{|m|}(k_{0 \perp  m n}\rho)$ (the index must be
non-negative to ensure the regularity at the origin). In this case
the eigenvalues $k_{0 \perp m n}$ can be determined by the zeros of the
Bessel functions. \\

According to \eq{effact}, the ghost contribution to the transverse
energy of the vortex is the sum of logarithms of the eigenvalues:
\begin{eqnarray}
\nonumber
{\cal E}_{{\rm ghost}}&=&-\sum_{c,m,n,k_\parallel}
\left[\ln E^2(c,m,n,k_\parallel)-\ln E_0^2(c,m,n,k_\parallel)\right]\\
\label{Egh1}
&=&-\sum_{c=1}^{N^2-1} \sum_{m=-\infty}^\infty \sum_n^\infty
\!\int\!\frac{d^{d-2}k_\parallel}{(2\pi)^{d-2}}
\left[\ln\left(k_\parallel^2 + k_{\perp\,cmn}^2\right)
-\ln \left(k_\parallel^2+ k_{0\perp\,mn}^2\right)\right]\;,
\end{eqnarray}
where $k_{0\perp\,mn}^2$ are the eigenvalues of the free operator in a box,
with $\mu(\rho)$ set to zero. Color polarizations with $\alpha_c=0$ do not
contribute since at $\alpha_c=0$ \eq{eig1} reduces to the free one. \\

We first integrate over $k_\parallel$. In the $3d$ case the integration
is finite:

\begin{equation}
\int_{-\infty}^\infty\frac{dk_\parallel}{2\pi}
\ln \left(\frac{k_\parallel^2+k_\perp^2 }{k_\parallel^2+k_{0 \perp}^2}
\right) =  |k_\perp|-|k_{0\perp }|.
\label{kpar3}\end{equation}
In the $4d$ case integration over $k_\parallel$ needs to be regularized.
We use the Pauli--Villars regularization scheme, so that henceforth
$\Lambda_{{\rm QCD}}=\Lambda_{P.V.}=e^{1/12}\Lambda_{\overline{MS}}$.
We have

\begin{equation}
\int\!\frac{d^2k_\parallel}{(2\pi)^2}
\ln\frac{(k_\parallel^2+k_\perp^2)(k_\parallel^2+k_{0\perp}^2+M^2)}
{(k_\parallel^2+k_{0\perp}^2)(k_\parallel^2+k_\perp^2+M^2)}
=\frac{1}{4\pi}\left[k_\perp^2\left(\ln\frac{M^2}{k_\perp^2}+1\right)
-k_{0\perp}^2\left(\ln\frac{M^2}{k_{0\perp}^2}+1\right)\right].
\label{kpar4}\end{equation}

There is a general formula for summation of any function of eigenvalues
becoming continuous in the limit when the box radius goes to
infinity, see Section 6. In the continuum limit the summation can be
replaced by integration over the spectrum with a weight being
the phase shift $\delta(k_\perp)$ of the corresponding differential
equation, in this case \eq{eig1}:

\begin{equation}
\sum_n\left[F(k_n)-F(k_{0n})\right]
= -\frac{1}{\pi}\int_0^\infty\!dk\,\delta(k)\frac{dF(k)}{dk}
\label{genphase}\end{equation}
(the subscript $\perp$ will be henceforth suppressed).
We shall describe the method of finding the phase shifts in Section
6. The phase shifts $\delta(k)$ depend on the `partial wave' $m$ and
on the color polarization $c$. Let us introduce the accumulated phase
shift, summing over all partial waves,

\begin{equation}
\delta_c(k)=2\sum_{m=1}^\infty \delta_{cm}(k) + \delta_{c0}(k).
\label{accphase}\end{equation}
Since $\alpha_c$ assumes as many ``plus'' values as there are ``minus''
ones, it is sufficient to sum over non-negative values of $m$.
[Alternatively, one can take only positive values of $\alpha_c$ but then
sum over all $m$'s, positive and negative.] Using the linearity of
\eq{genphase} in $\delta(k)$ one can introduce the aggregate phase shift
of the ghost operator,

\begin{equation}
\delta_{{\rm ghost}}(k)=\sum_c\delta_c(k).
\label{deltaghost}\end{equation}
It is a common function both for $3d$ and $4d$. However, the
expression for the ghost energy is different in $3d$ and $4d$, as the
weights with which eigenvalues are taken are different, cf.
\eqs{kpar3}{kpar4}. Using the general \eq{genphase} we obtain the
needed transverse energy of the vortex as due to the ghost part of
the quantum action :
\begin{eqnarray}
\label{psd3gh} {\cal E}_{{\rm ghost}}^{3d}
&=&\frac{1}{\pi}\int_0^\infty\!dk\,\delta_{{\rm ghost}}(k)\\
\label{psd4gh}
{\cal E}_{{\rm ghost}}^{4d}&=&\frac{1}{\pi^2}\int_0^\infty\!dk\,k\,
\ln\frac{M}{k}\,\delta_{{\rm ghost}}(k).
\end{eqnarray}

\subsection{The gluon operator}

We now consider the gluon operator $W_{\mu\nu}^{ab}$, see
\eq{quadrform}. Its $(D^2)^{ab}\delta_{\mu\nu}$ part is essentially
identical to the ghost operator. For the other part we find the
following eigenvalues:

\begin{eqnarray}
\label{eigbeta}
{\rm Eig} (F_{\mu\nu}^c f^{acb}) &=&
  \frac{\mu'(\rho)}{\rho}
 ( \{\alpha_c\}_{\rm color} \otimes \{\beta_\lambda\}_{\rm space}),
\qquad
\beta_\lambda =\left\{\begin{array}{cc}
 \{0, 1,-1\}  &  {\rm for \;\; 3d,}  \\
 & \\
\{0,0,1,-1\}  & {\rm for \;\; 4d}, \end{array} \right.
\end{eqnarray}
where the color eigenvalues $\alpha_c$ are the same as for the ghost
operator, see \eq{eigalpha}.

Therefore, the transverse eigenvalues $k_\perp$ of the gluon operator
are found from solving the differential equation for the
eigenfunctions $Z^{(c,\lambda)}_m(k_\perp\rho)$ where $c=1,...,(N_c^2-1)$
labels the color polarization and $\lambda=1,...,d$ labels the space
polarization of the gluon:

\begin{equation}
\left[-\frac{1}{\rho}\frac{\partial}{\partial \rho}
  \rho \frac{\partial}{\partial \rho}
+\frac{\left(m-\alpha_c\,\mu(\rho)\right)^2}{\rho^2}
-\frac{2\alpha_c\,\beta_\lambda\,\mu'(\rho)}{\rho}\right]
Z^{(c,\lambda)}_m(\rho)
=k_{\perp\,c\lambda m}^2 Z^{(c,\lambda)}_m(\rho).
\label{eig2}
\end{equation}

According to \eq{effact}, the gluon contribution to the transverse
energy of the vortex is the sum of logarithms of the eigenvalues:
\begin{equation}
{\cal E}_{{\rm gluon}}=\frac{1}{2}\sum_{c,\lambda,m,n,k_\parallel}
\left[\ln E^2(c,\lambda,m,n,k_\parallel)
-\ln E_0^2(c,\lambda,m,n,k_\parallel)\right].
\label{Egl1}\end{equation}

Color polarizations with $\alpha_c=0$ cancel from this equation.
In $4d$ there are two space polarizations with $\beta_\lambda=0$ for
which the eigenvalue equation \ur{eig2} reduces to that of the
ghost operator, \eq{eig1}. Therefore, ${\cal E}_{{\rm ghost}}$ is
completely canceled by that part of the gluon operator spectrum.
In $3d$ there is only one space polarization with $\beta_\lambda=0$,
so that the cancellation is not complete.\\

Similar to the previous subsection, we introduce the phase shifts
$\delta(k)$ but now corresponding to \eq{eig2}:

\begin{equation}
\delta_{{\rm gluon}}^{(d)}(k)=\sum_c\sum_\lambda\delta_{c\lambda}(k),
\qquad\delta_{c\lambda}(k)=2\sum_{m=1}^\infty \delta_{c\lambda m}(k)
+\delta_{c\lambda 0}(k).
\label{deltagluon}\end{equation}
Notice that, unlike the ghost case, here the aggregate phase shift
depends on the number of dimensions as the number of gluon
polarizations $\lambda$ is $d$-dependent. Using the results of the
$k_\parallel$ integration from the previous subsection we can write
down the gluon contribution to the transverse energy via the
aggregate phase shifts:
\begin{eqnarray}
\label{E3gluon}
{\cal E}_{{\rm gluon}}^{3d}&=&
-\frac{1}{2\pi}\int_0^\infty\!dk\,\delta_{{\rm gluon}}^{(3)}(k),\\
\label{E4gluon}
{\cal E}_{{\rm gluon}}^{4d}&=&
-\frac{1}{2\pi^2} \int_0^\infty\!dk\,k\,\ln\frac{M}{k}\,
\delta^{(4)}_{{\rm gluon}}(k).
\end{eqnarray}

\subsection{Full quantum energy}

Adding up the ghost and gluon contributions to the quantum energy we get:
\begin{eqnarray}
\label{full3} {\cal E}_{\rm quant}^{3d}
&=&\frac{1}{\pi}\int_0^\infty\!dk\,
\left[
\delta_{{\rm ghost}}(k)
-
\half\delta_{{\rm gluon}}^{(3)}(k)
\right],\\
\label{full4}
{\cal E}_{\rm quant}^{4d}
&=&\frac{1}{\pi^2}\int_0^\infty\!dk\,k\,
\ln\frac{M}{k}\,
\left[
\delta_{{\rm ghost}}(k)
-
\half\delta_{{\rm gluon}}^{(4)}(k)
\right].
\end{eqnarray}
The combination of ghost and gluon phase shifts in the square brackets will be
called the full phase shift; for various cases they are plotted in Figs. 2,4
and 5.

\subsection{Scaling properties}

Let as assume that we have calculated the ghost and gluon phase
shifts for a certain profile function $\mu(\rho/\rho_0)$ with
$\rho_0=1$ in some arbitrarily chosen units. If we now change
$\rho_0$ all quantities have to scale in accordance with the
dimensional analysis of Section 4. Equations for ${\cal E}_{\rm quant}^{3d,4d}$
of the previous subsection can be translated
into the following expressions for the constants $A,B,C$ introduced in
Section 4:

\begin{eqnarray}
\label{Aa}
A&=&\pi\rho_0^2\int_0^\infty\!\frac{d\rho}{\rho}(\mu^\prime(\rho))^2,\\
\nonumber \\
\label{C}
C&=&\frac{\rho_0}{\pi}\int_0^\infty\!dk\,
\left[
\frac{1}{2}\delta_{{\rm gluon}}^{(3)}(k)
-
\delta_{{\rm ghost}}(k)
\right],\\
\nonumber \\
\label{tildeA}
\tilde A &=& \frac{24\, \rho_0^2}{11\,N}\int_0^\infty\!dk\,k\,
\left[
\frac{1}{2}\delta_{{\rm gluon}}^{(4)}(k)
-
\delta_{{\rm ghost}}(k)
\right], \\
\nonumber \\
\label{B}
\ln B&=&-\frac{1}{\tilde A}\frac{24\,\rho_0^2}{11\,N}
\int_0^\infty\!dk\,k\,\ln (k\rho_0)\,
\left[
\frac{1}{2}\delta_{{\rm gluon}}^{(4)}(k)
-
\delta_{{\rm ghost}}(k)
\right].
\end{eqnarray}

As discussed in Section 4, the renormalizability of Yang--Mills
theory in $4d$ requires that $\tilde A=A$: if that is satisfied the
effective action does not depend on the ultraviolet cutoff $M$. This
relation is, therefore, an important check of the numerics.
It can be said that the calculation of $\tilde A$ is a particular way of
getting numerically the `11/3' of the asymptotic freedom law. Further on,
we verify the scaling relations \urs{e4de}{e3de} explicitly by
changing the scale $\rho_0$ of the trial functions and checking that
$\tilde A,B,C$ are $\rho_0$-independent (see also Fig.~\ref{scalingflux0}).  \\

Another powerful check of our numerical performance is provided by
the number of colors $N$: the quantities $\tilde A$  and $B$ should be
independent of $N$ while $C$ must be proportional to $N$.
The calculations successfully pass these tests for the $\tilde A$
and $C$ within small errors, whereas the sensitivity of $B$ on the
numerics is considerably larger,  see Table~\ref{resdetails}. \\

%%%%%%%%%%%%%%%%%%%%%%%%% Table 1  %%%%%%%%%%%%%%%%%%%%%%%%%%%%%%%%%%%%%%%%
\begin{table}
\begin{tabular}{|l||r|r||r|r||r|r|}
\hline
& \multicolumn{2}{|c||}{}
& \multicolumn{2}{|c||}{}
& \multicolumn{2}{|c|}{}\\
& \multicolumn{2}{|c||}{ $\mu(\rho) = \exp(-\rho_0/\rho-\rho/\rho_0)$}
& \multicolumn{2}{|c||}{ $\mu(\rho) = \exp(-(\rho_0/\rho)^3)$}
& \multicolumn{2}{|c|}{ $\mu(\rho) = (\rho/\rho_0)^6/((\rho/\rho_0)^6+1)$} \\
& \multicolumn{2}{|c||}{}
& \multicolumn{2}{|c||}{}
& \multicolumn{2}{|c|}{}\\
\hline
&&&&&&\\
& $\rho_0 = 1$ & $\rho_0 = \sqrt{2}$
& $\rho_0 = 1$ & $\rho_0 = \sqrt{2}$
& $\rho_0 = 1$ & $\rho_0 = \sqrt{2}$ \\
&&&&&& \\
\hline
\hline
&&&&&&\\
$A$              &  0.2424 &  0.2424 &  2.2333 & 2.2333 & 3.3767 & 3.3767\\
$N_c =2$: &&&&&& \\&&&&&& \\
$A/\tilde A$     & 1.0603& 1.0066  &1.0007 & 0.9862 & 1.0155 &  1.0026\\
$B$              & 1.1686 & 1.0692  & 1.0388  & 1.0204  &  0.8854 &  0.8923   \\
$C$              & 0.1094 & 0.1089 & 1.1651 & 1.1513 &  1.5025 &  1.4693  \\ &&&&&& \\
&&&&&& \\
\hline
&&&&&& \\
$N_c =3$: &&&&&& \\ &&&&&& \\
$\frac{\tilde A(N_c=2)}{\tilde A(N_c=3)}$ & 1.0000 & 0.9999 & 1.0293 & 1.0331 & 1.0107  &  1.0193 \\
$\frac{B(N_c=2)}{B(N_c=3)}$               & 1.0082 & 1.0075 & 0.8767 & 0.8750 & 0.8751  &  0.8747  \\
$\frac{C(N_c=2)}{C(N_c=3)}$               & 0.6774 & 0.6766 & 0.6484 & 0.6511 & 0.6388  &  0.6387 \\ &&&&&& \\
\hline
\end{tabular}
\caption{Results for flux-0 and flux-1 profiles used.
The table shows two columns for each profile one for $\rho_0=1$ and one
for $\rho_0=\sqrt{2}$.}
\label{resdetails}
\end{table}
%%%%%%%%%%%%%%%%%%%%%%%%%%%%%%%%%%%%%%%%%%%%%%%%%%%%%%%%%%%%%%%%%%%%%%%%%

\section{The phase shift method}
\label{secphsm}

The $Z(2)$ vortex field $\mu(\rho)$ with $\mu(0)=0$ and
$\mu(\infty)=1$ is very specific from the point of view of
differential \eqs{eig1}{eig2}. Near the origin $\rho\approx 0$ the
equations resemble the free equation for the Bessel functions
$J_m(k\rho)$ while at infinity $\rho\to\infty$ they resemble
the equation for $J_{m\pm 1}(k\rho)$, with a shifted index.
This is the essence of quantized $Z(2)$ vortices, see in this respect
ref. \cite{Diakonov:1999gg}. \\

This fundamental property of $Z(2)$ vortices causes certain
difficulties in finding the continuous spectrum of eigenvalues in a
given $m$ sector. We present here an economical method of dealing
with this problem; it enables one to find the spectrum unambiguously.
We have borrowed the idea of the method from Jost scattering theory,
generalizing it to problems with cylindrical symmetry.  The
application of the phase-shift method to 1-loop calculations
of the effective action has been, in general terms, developed
in Refs. \cite{GJQW,DPP}. \\

In the following we consider a general differential equation

\begin{equation}
\left[-\frac{1}{\rho}\frac{\partial}{\partial\rho}
\rho\frac{\partial}{\partial\rho}
+\frac{m^2}{\rho^2} + U(\rho)\right] Z(\rho)=k^2 Z(\rho),
\label{diffeq}\end{equation}
the equation for ghosts \ur{eig1} and for gluons \ur{eig2}
being exactly of that form. \\

One looks for the solution which is regular at the origin and follows
the oscillatory asymptotics of the Bessel function at infinity:

\begin{equation}
Z(r)=\sqrt{\frac{2}{\pi k\rho}}
\cos\left[k\rho-\frac{\pi}{4}-\frac{m\pi}{2}+\delta(k)\right].
\label{asympt}\end{equation}
This serves a definition of the phase shift $\delta(k)$. If the
system is put in the box of large radius $R$ the zeros of \ur{asympt}
define the spectrum:

\begin{equation}
k_n=k_{0n}-\frac{\delta(k_n)}{R},\qquad
k_{0n}=\frac{\pi}{R}\left(n+\frac{3}{4}+\frac{m}{2}\right)
\label{spectrum}\end{equation}
where $k_{0n}$ is the spectrum with zero potential. A sum of any
function of the eigenvalues $k_n$ in the continuum limit
($R\to \infty $) becomes:

\begin{equation}
\sum_n\left[F(k_n)-F(k_{0n})\right]=-\frac{1}{R}\sum_n F^\prime(k_n)
\delta(k_n)=-\frac{1}{\pi}\int_0^\infty\!dk\, F^\prime(k)\delta(k).
\label{genphase1}\end{equation}
This is \eq{genphase} used in the previous section. \\

In the free case ($U=0$) the differential \eq{diffeq} have
regular solutions $J_m(k\rho)$ (the ordinary Bessel functions),
as well as singular solutions $Y_m(k\rho)$ (v.Neumann functions).
From those one combines Hankel functions,

\begin{equation}
H^{\pm}_m(k\rho) = J_m(k\rho) \pm i Y_m(k\rho).
\label{Hankel}\end{equation}
Taking into account that $J_m$ and $Y_m$ are both solutions of the
same differential equation, their Wronskian is a constant:

\begin{equation}
W(J_m,Y_m) \equiv k\rho\left[J_m(k\rho) Y_m'(k\rho)
-J_m'(k\rho) Y_m(k\rho)\right] = \frac{2}{\pi}.
\label{Wronsk}\end{equation}
Using \ur{Wronsk} one can easily verify that the needed solution
$Z_m(k\rho)$ of \eq{diffeq} satisfies the integral equation:

\begin{eqnarray}
\nonumber
Z_m(k\rho) &=& J_m(k\rho) - \frac{\pi}{2}
\int_0^\rho d\rho'\rho' \left[J_m(k\rho) Y_m(k\rho')- J_m(k\rho')
Y_m(k\rho)\right] U(\rho') Z_m(k\rho')\\
\nonumber
&=& {\rm Im} \left\{i H^{(+)}_m(k\rho) \left[ 1-\frac{i\pi}{2}
\int_0^\rho d\rho'\rho' H^{(-)}_m(k\rho') U(\rho')
Z_m(k\rho')\right]\right\} \\
\label{identity}
&=& {\rm Re} \left\{ H^{(+)}_m(k\rho) \left[ 1-\frac{i\pi}{2}
\int_0^\rho d\rho'\rho'
H^{(-)}_m(k\rho') U(\rho') Z_m(k\rho')\right]\right\}.
\end{eqnarray}
It enables one to find $Z_m(k\rho)$ iteratively knowing the function
at smaller values of $\rho$, starting from $\rho=0$. \\

At $\rho\to\infty$ one obtains the asymptotics:

\begin{equation}
A \cos\!\left[k\rho\!-\!\frac{\pi}{4}\!-\!\frac{m\pi}{2}\!+\!\delta(k)\right]
={\rm Re}\left(e^{i(k\rho-\pi m/2-\pi/4)}\!\left[1\!-\!\frac{i\pi}{2}
\!\int_0^\infty\!\!d\rho'\rho'
H^{(-)}_m(k\rho') U(\rho') Z_m(k\rho')\right]\right).
\label{Jost1}\end{equation}
Putting $k\rho-m\pi/2 -\pi/4=2n\pi$ or $k\rho-m\pi/2 -\pi/4=(2n+1/2)\pi$
with $n$ being a large integer and comparing the expressions we obtain the
equation for the phase shift:

\begin{equation}
\tan \delta(k)=\frac{{\rm Im} f_m(k)}{{\rm Re} f_m(k)},\qquad
f_m(k)=1-\frac{i\pi}{2}
\int_0^\infty d\rho \rho H_m^{(-)}(k \rho) U(\rho) Z_m(k\rho)\;.
\label{josteq}\end{equation}
We shall refer to \eq{josteq} as Jost equation as it is closely related
to the theory of Jost functions developed in scattering theory,
see e.g. \cite{Taylor}. \\

\begin{figure}
\centerline{\psfig{figure=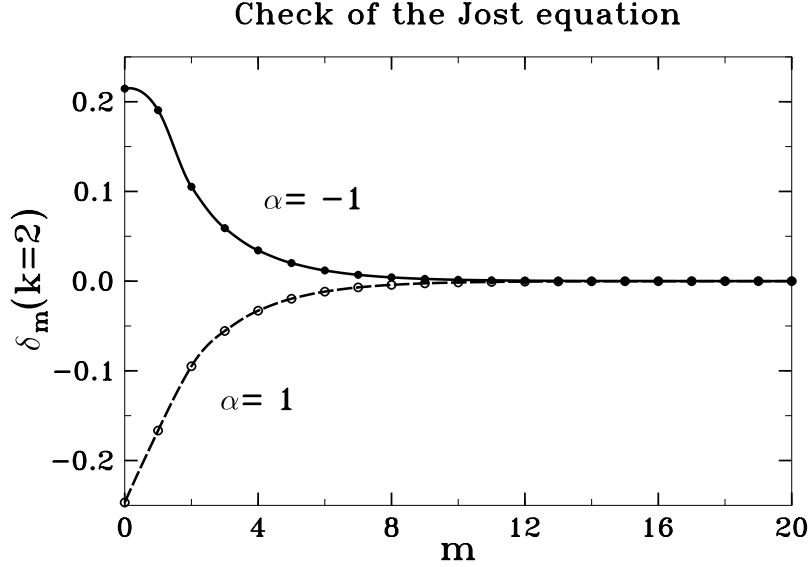,width=13cm}}
\caption{Check of \eq{josteq}. For $k=2$, $\alpha=\pm 1$ and
$\beta=1$ we show the phase shift as function of $m$. The solid and
dashed line present the phase shifts as obtained by extrapolating
the difference of the zeros, while the open and filled circles show
the result as obtained from \eq{josteq}. One finds perfect
agreement between the two methods. It is furthermore seen that the
phase shifts decay rapidly in $m$.}
\label{checkjost}
\end{figure}

We performed a numerical check of \eq{josteq} by taking, as an example,
the potential arising from the gluon operator (see \eq{eig2}),

\begin{equation}
U(\rho)=\frac{1}{\rho^2} [(m-\alpha \mu(\rho))^2-m^2 -
2\alpha\beta \rho \mu'(\rho)],\quad{\rm with}\quad
\mu(\rho)=\exp(-\rho-1/\rho).
\label{U1}\end{equation}
We computed the phase shifts directly from the shift of zeros of
$Z_m(k\rho)$ with respect to those of the free $J_m(k\rho)$, and then
compared with phase shifts obtained from \eq{josteq}. The comparison
of the two methods is shown in Fig.~\ref{checkjost}: the agreement
is excellent. It is also seen that $\delta_m(k)$ decays fast as
function of $m$.

%%%%%%%%%  FIGURE 1 %%%%%%%%%%%%%%%
%%%%%%%%%%%%%%%%%%%%%%%%%%%%%%%%%%%%

One can simplify the method of finding the phase shifts even further.
The second integration in \eq{josteq} is in fact unnecessary. Indeed,
from \eq{identity} one has the identity:

\begin{equation}
Z_m(k\rho)=\frac{1}{2}\Bigg[H_m^{(+)}(k\rho) f_m(k)
+ H_m^{(-)}(k\rho) f^*_m(k) \Bigg].
\label{id2}\end{equation}
Putting it in the Wronskian (see the definition \ur{Wronsk}) we get:

\begin{equation}
W(Z_m,H_m^{(-)})=\frac{1}{2} f_m(k)\;W(H_m^{(+)},H_m^{(-)})
=i f_m(k)\; W(Y_m,J_m)=-\frac{2i}{\pi} f_m(k).
\label{Wronsk2}\end{equation}
Therefore the needed phase shift can be found directly from the Wronskian
composed from the solution $Z_m(k\rho)$ and the Hankel function:

\begin{equation}
\delta_m (k)={\rm Arg}(f_m(k))
={\rm Arg}\left[\frac{i\pi}{2}W\left(Z_m,H_{|m|}^{(-)}\right)\right].
\label{dw}\end{equation}
Here ${\rm Arg}(re^{i\delta})=\delta$ is the argument of a complex
number. Notice that \eq{dw} is independent of $\rho$. Therefore,
evaluating the Wronskian in eq.~\ur{Wronsk2} at different values of $\rho$
is a valuable check  of the stability of the numerics. In fact,
this method proves to be quite stable; we actually use it to calculate
the phase shifts.

%%%%%%%%%%%%%%%%%%%%%%%%% Figure 2 %%%%%%%%%%%%%%%%%%%%%%%%%%%%%%%%%%%%%%%%
\begin{figure}
\centerline{\psfig{figure=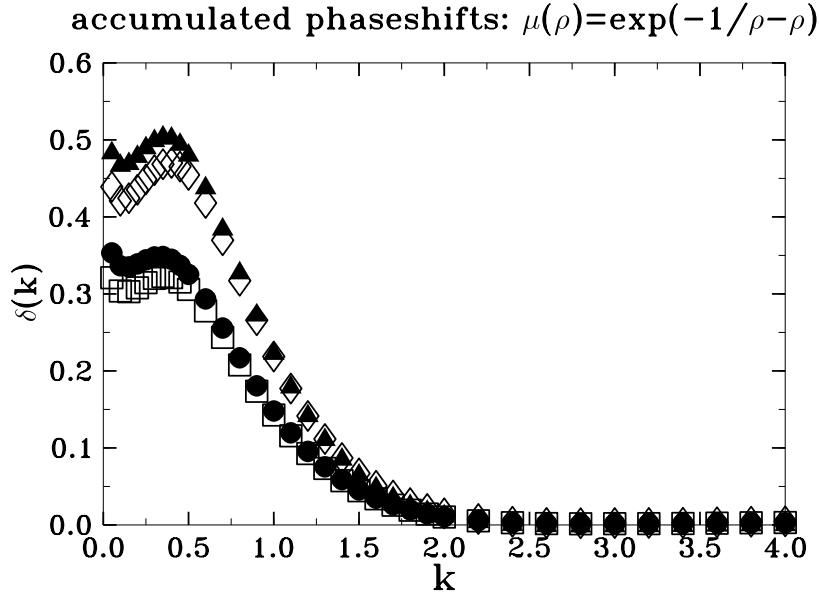,width=13cm}}
\caption{
Full phase shifts $\delta = \half\delta_{\rm gluon}-\delta_{\rm  ghost}$
for the profile $\mu(\rho) = \exp(-1/\rho-\rho)$:
 $\delta_{N=2}^{(3d)}(k)$ (filled circles),
 $\delta_{N=3}^{(3d)}(k)$ (filled triangles),
 $\delta_{N=2}^{(4d)}(k)$ (open squares),
 $\delta_{N=3}^{(4d)}(k)$ (open diamonds).
 }
\label{aphflux0}
\end{figure}
%%%%%%%%%%%%%%%%%%%%%%%%%%%%%%%%%%%%%%%%%%%%%%%%%%%%%%%%%%%%%%%%%%%%%%%%%%%
%
%
\begin{figure}
\centerline{\psfig{figure=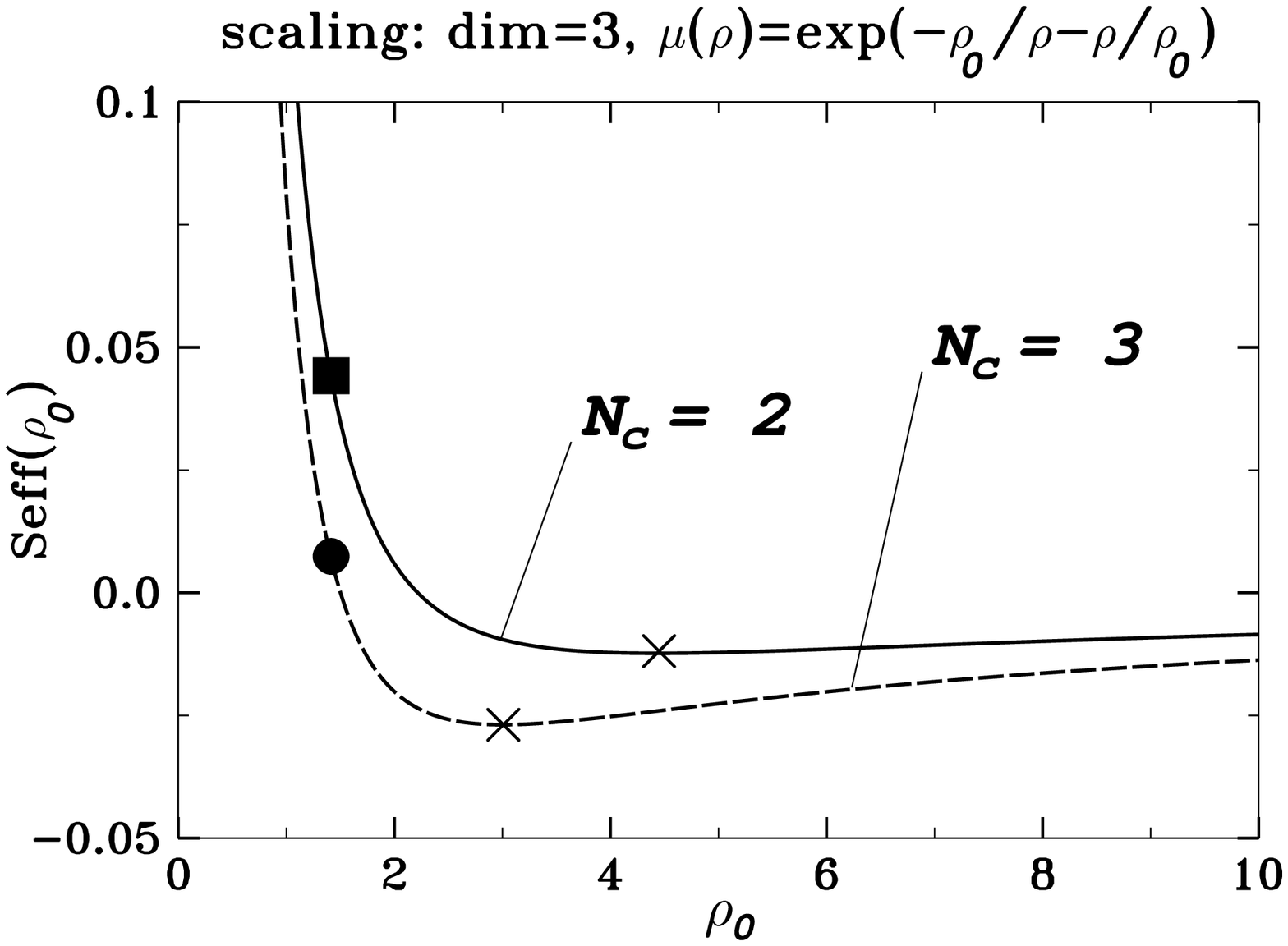,width=14cm}}
\centerline{\psfig{figure=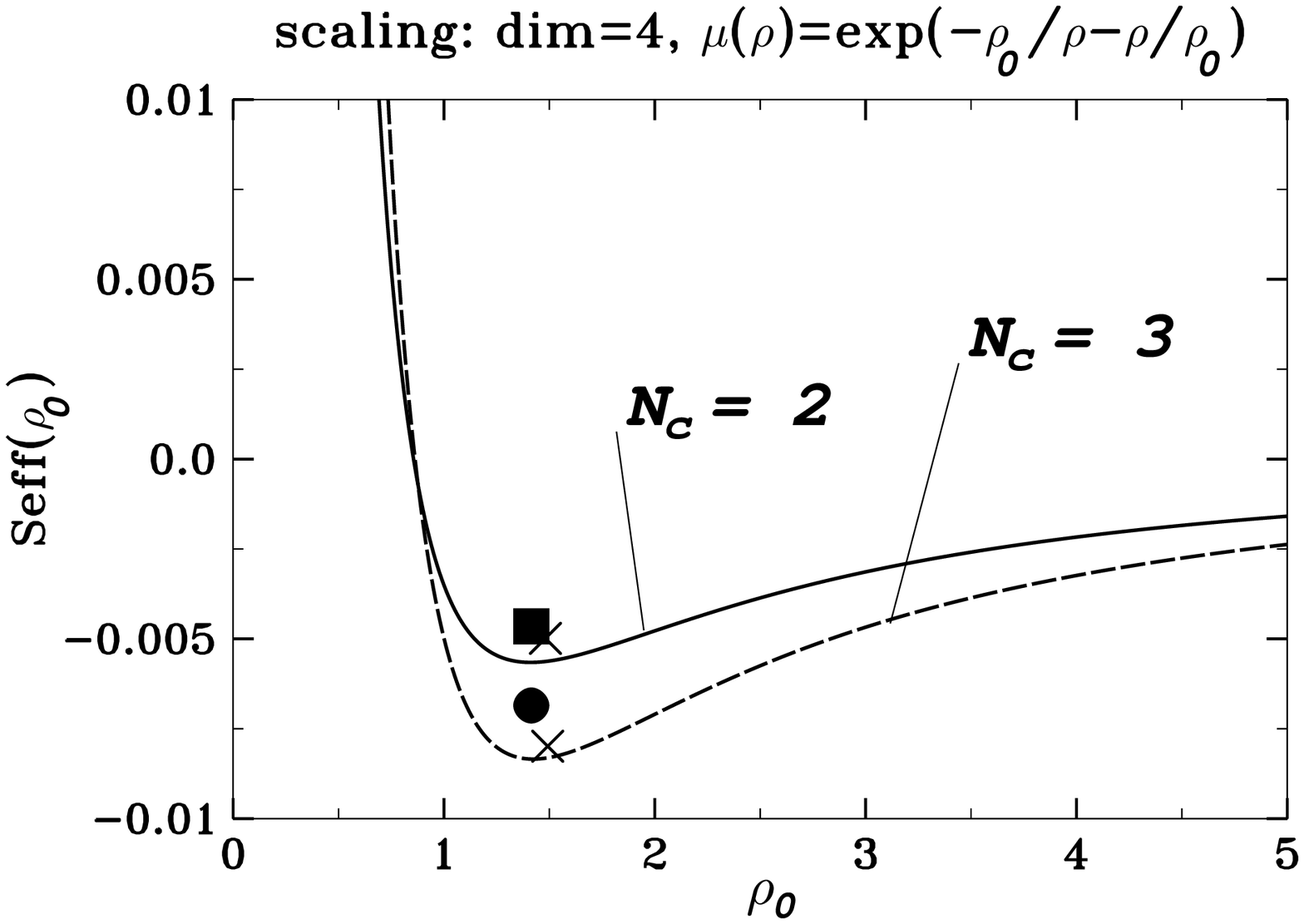,width=14cm}}
\caption{Scaling behavior of the effective action in three and four
dimensions: The figures show the effective action in three and four
dimensions for the profile $\mu(\rho) = \exp(-\rho_0/\rho-\rho/\rho_0)$ as a
function of $\rho_0$. The solid line shows the solution for $N_c=2$ and the
dashed line for $N_c=3$. The dimensionfull constants have been set to 1, i.e.
$g_3=1$ in three dimensions and $\Lambda_{\rm QCD}=1$ in four dimensions.
The filled circle and the square come from an independent calculation
with $\rho_0=\sqrt{2}$ explicitly plugged in. The crosses denote the place
of the minimum of the effective action averaged over the results
for $\rho_0=1$ and $\rho_0=\sqrt{2}$, c.f. Table~\ref{sumres}.}
\label{scalingflux0} \end{figure}

\section{Results}
\label{secnum}
\subsection{Flux-0 vortices}

We first consider not a $Z(2)$ but a zero-flux vortex with
$\mu(0)=\mu(\infty)=0$. For definiteness we take a smooth profile
of the form:

\begin{equation}
\mu(\rho) = \exp\left(-\rho-\frac{1}{\rho}\right).
\label{mu0}\end{equation}
Let us briefly describe the numerical procedure. We start by defining
a range $R$ for a given value of $k$ and $m$ up to which the differential
equations have to be solved. To make sure that we are working in an
asymptotic regime we choose $R_{\rm max}= {\rm max}(60.0,\,x_{|m|15})/k$
where $x_{|m|n}$ is the $n$-th zero of the Bessel function $J_{|m|}(x)$.
We have checked that for this choice the phase shifts calculated from
\eq{josteq} and from \eq{dw} do agree so that we are indeed reaching
the asymptotic regime. \\

The initial condition for finding a regular solution $Z_m(\rho)$
is the free Bessel function which behaves as $J_m(k\rho)\approx
(k\rho/2)^m/\Gamma(m+1)$ at small $\rho$: it is very small at large $m$.
Therefore, for small $\rho$ (actually starting from some finite
but small $\rho_{\rm min} = 0.00001/k$ and for $m\ge 4$) we solve,
instead of \eq{diffeq}, the corresponding equation for the function
$B(\rho) = Z(\rho)/((\rho/2)^m/\Gamma(m+1))$,

\begin{equation}
B''(\rho) = -B'(\rho)\frac{2m + 1}{\rho} +
 \frac{B(\rho)((m - \alpha\mu(\rho))^2 - m^2
- 2 \alpha \beta \rho \mu'(\rho))}{\rho^2} - k^2B(\rho)\;.
\label{Beq}\end{equation}
with the initial conditions

\begin{equation}
B(\rho_{\rm min})=1,\qquad
B'(\rho_{\rm min}) = -k\frac{k\rho_{\rm min}}{2(m + 1)}
\left(1 - \frac{(k\rho_{\rm min})^2}{4(m + 2)}\right).
\label{initcond}
\end{equation}
This differential equation is solved up to the point $R_p = 0.9
x_{|m|1}/k$ where $x_{|m|1}$ is the first zero of the corresponding
Bessel function. Starting from this point the original differential
\eq{diffeq} is solved, with the initial conditions:

\begin{eqnarray}
\nonumber
Z(R_p) &=& \frac{B(R_p)}{\Gamma(m+1)}
\left(\frac{kR_p}{2}\right)^m \\
\nonumber
Z^\prime(R_p) &=&
\frac{B'(R_p)}{\Gamma(m+1)} \left(\frac{kR_p}{2}\right)^m +
\frac{k B(R_p)}{2\Gamma(m)} \left(\frac{kR_p}{2}\right)^{m-1}\,.
\end{eqnarray}
We have checked that this procedure reproduces the usual Bessel functions
in the free case, when $\mu(\rho)=0$. \\

Having found the regular solution $Z_m(\rho)$, we use it to calculate
phase shifts in three different ways. The first method is to use
\eq{josteq}, where the integration goes up to $R_{\rm max}$.
This procedure becomes sometimes unstable because for large $\rho$ the
integrand is oscillating. The two other methods exploit \eq{dw}. We
evaluate the Wronskian either at $0.8R_{{\rm max}}$ or at $0.9R_{\rm max}$.
We have checked that except in cases where the phase shift is consistent
with zero these three different methods agree with each other. The phase
shifts actually used in the results are the ones taken from the
Wronskian evaluated at $0.8R_{\rm max}$. \\

The individual phase shifts at given $k$ are then accumulated over
$m$, $c$ and $\lambda$. Fig.~\ref{aphflux0} shows the total aggregate
phase shifts for N=2,3 in three and four dimensions. It is seen
that in all cases $\delta(k)$ goes to zero quite fast for large $k$.
For very small $k$ values $\delta(k)$ rises again after a maximum at
$k=0.4\;$. \\

The results are shown in Table~\ref{resdetails}. The $A$ coefficients of the
classical and the quantum parts are equal within a small error due to
the numerics, as it should be in order to ensure that
in four dimensions the results are independent of the UV cutoff $M$.
It is also seen that the $A$ coefficients are independent of the number of
colors. \\

As a final step we check the scaling behavior predicted by
\eqs{e4de}{e3de}. To that end we perform
the same calculation again but for a profile $\mu(\rho) =
\exp(-\sqrt{2}/\rho-\rho/\sqrt{2})$ and compare the result with
the one obtained from the profile $\mu(\rho)=\exp(-1/\rho-\rho)$ and
assuming the scaling \eqs{e4de}{e3de}.
Tab.~\ref{resdetails} shows indeed that in $3d$ and $4d$
the scaling law is fulfilled, which is yet another important cross
check of our calculation.  Furthermore, it is seen that in all cases
we find a minimum of the effective action at negative values, which
indicates that the perturbative vacuum is unstable against the
production of zero-flux vortices. The values for the minima of the scaling
factor are given in Table~\ref{sumres}.

\subsection {Flux-1 vortices}

For the calculations with the $Z(2)$ unit-flux vortices we use the
following trick which enables us to use the same technique as
used for the flux-0 vortices. Instead of the actual flux-1 profile
$\mu(\rho)$ with $\mu(0)=0$ and $\mu(\infty)=1$, we consider the
profile $1-\mu(\rho)$. From the energy point of view nothing is
changed since the classical part depends only on the square of the
first derivative of $\mu(\rho)$ while in the quantum part it
corresponds to an overall shift of the summation in $m$. We have checked
that in case of SU(2) this `reflection' indeed produces
the same results as if one makes the calculation directly with the flux-1
profile  $\mu(\rho)$. For the $SU(2)$ vortex embedding into SU(3)
this is not the case
because here we have a contribution corresponding to the eigenvalues 1/2 in
color space and the shift in integer units in the summation over $m$ does
not work. \\

The numerics for the flux-1 vortices is more
difficult than for the flux-0 ones.
The first problem arises due to the fact that the phase shift exceeds
$\pi$ in magnitude. As the arcus tangent  is a multivalued function
one then has to reconstruct the true function $\delta(k)$ by adding
suitable multiples of $2\pi$ so that the resulting function is
continuous in $k$. Furthermore, general constants in terms of suitable
multiples of $\pi/4$ have to be added so that the aggregate phase
shifts match each other for $k\to \infty$ at a value close to zero.\\

\begin{figure}
\centerline{\psfig{figure=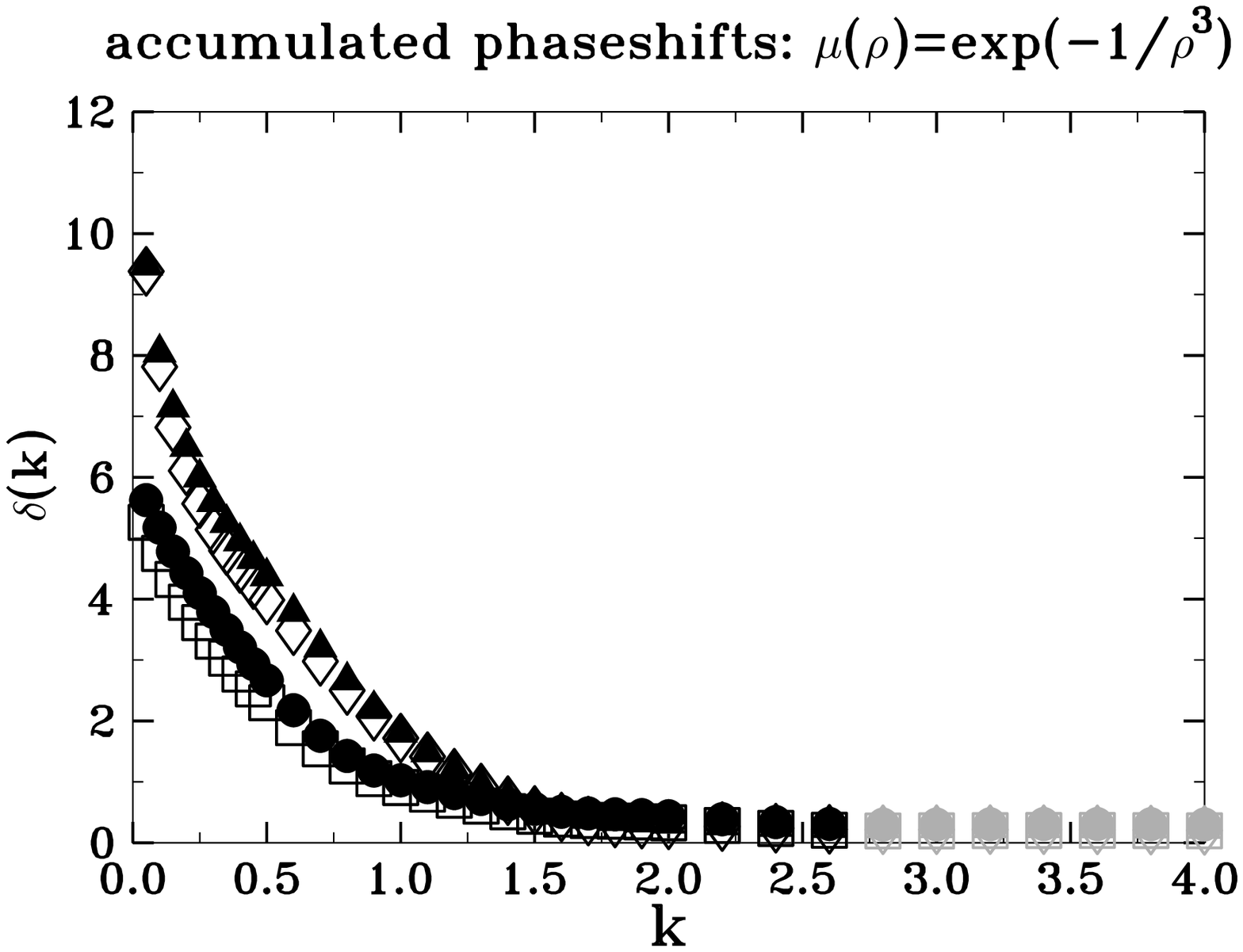,width=13cm}}
\caption{
Full phase shifts $\delta = \half\delta_{\rm gluon}-\delta_{\rm ghost}$
for the profile $\mu(\rho) = \exp(-1/r^3)$:
 $\delta_{N=2}^{(3d)}(k)$ (filled circles),
 $\delta_{N=3}^{(3d)}(k)$ (filled triangles),
 $\delta_{N=2}^{(4d)}(k)$ (open squares),
 $\delta_{N=3}^{(4d)}(k)$ (open diamonds).
 The points in grey have been excluded from the analysis as the
function $\delta(k)$ goes over to a spurious constant behavior.}
\label{inve3}
\end{figure}

\begin{figure}
\centerline{\psfig{figure=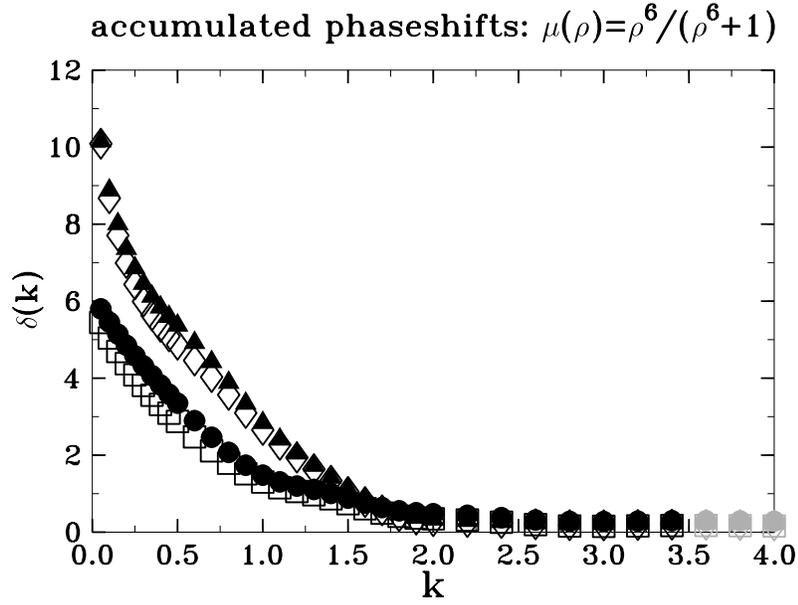,width=13cm}}
\caption{
Full  phase shifts $\delta = \half\delta_{\rm gluon}-\delta_{\rm ghost}$
for the profile $\mu(\rho) = \rho^6/(\rho^6+1)$:
 $\delta_{N=2}^{(3d)}(k)$ (filled circles),
 $\delta_{N=3}^{(3d)}(k)$ (filled triangles),
 $\delta_{N=2}^{(4d)}(k)$ (open squares),
 $\delta_{N=3}^{(4d)}(k)$ (open diamonds).
The points in grey have been excluded from the analysis as the
function $\delta(k)$ goes over to a spurious constant behavior.}
\label{invp6}
\end{figure}

The second technical problem is that, literally speaking,
$\delta(k)$ does not go to
zero with $k\to \infty$, but rather approaches a small constant, see
Figs. ~\ref{inve3},~\ref{invp6}. However, it seems to be a numerical
artefact since, as one
lowers  $\rho_{\rm min}$ in eq.~(\ref{initcond}), the small constant becomes
even smaller. Unfortunately, the numerical precision
does not allow a complete annihilation of this spurious behavior.
In fact, we have to make a cut at the point in k when
the function becomes a (tiny) constant.
The cross checks of renormalizability and the $N$ dependence are then
fulfilled. We obtain for two different flux-1 profiles,
\begin{eqnarray}
\mu^{(1)}(\rho) &=&
\exp(-1/\rho^3), \nonumber \\
\mu^{(2)}(\rho) &=& \rho^6/(\rho^6+1),
\end{eqnarray}
the values of the coefficients in the effective action as given in
Table~\ref{resdetails} and Table~\ref{sumres}. The cross checks
of renormalizability and N-dependence are fulfilled up to 5\% as
seen from Table~\ref{resdetails}. Again in all cases the effective action
exhibits a minimum.
In all cases the minima are at negative values of the effective
action, which means that the perturbative YM vacuum is unstable against the
production of center vortices in the case of SU(2) in three and four
dimensions and that it is even unstable with respect to the production of the
embedding of center vortices in SU(3). Comparing the two profiles used
it is seen that the power-like profile $\mu^{(2)}$ gives a slightly deeper
minima than the exponential-like profile $\mu^{(1)}$.
%
%

%%%%%%%%%%%%%%%%%%%%%%%% Table 2 %%%%%%%%%%%%%%%%%%%%%%%%%%%%%%%%%%%%%%%%
\begin{table}
\begin{tabular}{|l||r|r|r|}
\hline
&&& \\
&  $\mu(\rho) = \exp(-\rho_0/\rho-\rho/\rho_0)$
& $\mu(\rho) = \exp(-(\rho_0/\rho)^3)$
& $\mu(\rho) = (\rho/\rho_0)^6/((\rho/\rho_0)^6+1)$\\
&&& \\
\hline
&&& \\
$N_c =2$: &&&\\ &&& \\
$\rho_{0\,\rm min}^{(3d)}\;[g_3^{-2}]$ &  4.441$\pm$ 0.014  &
3.857$\pm$ 0.032 & 4.546$\pm$ 0.072 \\
${\cal E   }_{\rm min}^{(3d)}\;[g_3^{2}]$ &-0.012$\pm$ 0.000  &-0.150$\pm$ 0.003 &-0.163$\pm$ 0.005 \\
&&& \\
$\rho_{0\,\rm min}^{(4d)}\;[\Lambda^{-1}]$ &  1.477 $\pm$ 0.093   &
1.601 $\pm$ 0.020   &1.855 $\pm$0.010   \\
${\cal E   }_{\rm min}^{(4d)}\;[\Lambda^{2}]$  & -0.005 $\pm$ 0.001   & -0.040 $\pm$ 0.001   &-0.046 $\pm$ 0.000  \\
&&& \\
&&& \\
&&& \\ \hline
&&& \\
$N_c =3$: &&&\\ &&& \\
$\rho_{0\,\rm min}^{(3d)}\;[g_3^{-2}]$ & 3.007$\pm$ 0.007 &
2.506$\pm$ 0.028 &  2.903$\pm$ 0.046 \\
${\cal E   }_{\rm min}^{(3d)}\;[g_3^{2}]$ &-0.027$\pm$ 0.000 &-0.356$\pm$ 0.008 & -0.401$\pm$ 0.013 \\
&&& \\
$\rho_{0\,\rm min}^{(3d)}\;[\Lambda^{-1}]$ &1.488  $\pm$  0.093   &
1.403 $\pm$ 0.016   & 1.623  $\pm$ 0.009 \\
${\cal E   }_{\rm min}^{(3d)}\;[\Lambda^{2}]$  &-0.008  $\pm$  0.001  &
-0.079 $\pm$0.002   & -0.089  $\pm$ 0.001   \\
&&& \\
\hline
\end{tabular}
\caption{Summary of the physical results with error bars. The physical result
is the mean value from the two calculations at $\rho_0 =1,\sqrt{2}$, while
the error is given by the difference of these two values divided by the square
root of two.}
\label{sumres}
\end{table}
%
%
%%%%%%%%%%%%%%%%%%%%%%%%% Figure 3 %%%%%%%%%%%%%%%%%%%%%%%%%%%%%%%%%%%%%%%%

%%%%%%%%%%%%%%%%%%%%%%%%%%%%%%%%%%%%%%%%%%%%%%%%%%%%%%%%%%%%%%%%%%%%%%%%%%%

%%%%%%%%%%%%%%%%%%%%%%%%% Figure 4 %%%%%%%%%%%%%%%%%%%%%%%%%%%%%%%%%%%%%%%
%%%%%%%%%%%%%%%%%%%%%%%%%%%%%%%%%%%%%%%%%%%%%%%%%%%%%%%%%%%%%%%%%%%%%%%%%%%

\section{Summary and outlook}

We have calculated the energy of quantum fluctuations (gluon and
ghost) about vortices of various profiles, both in 4 and 3 Euclidean
dimensions. Center vortices are peculiar in that the level
counting for quantum fluctuations as compared to the free case is
not apparent. Because of this, the mere finiteness of the quantum energy
in the center-vortex background is not evident before an accurate
calculation is performed. We have developed an economical method
of `level counting' through phase shifts of the appropriate differential
operators, which is similar to that used in Jost's scattering theory,
and expressed the energy of quantum fluctuations through the
phase shifts. \\

Our results surpass many consistency tests. In $4d$ we actually
manage to calculate numerically and with a good accuracy the ``11/3''
of the asymptotic freedom law; as a result we get the `transmutation
of dimensions' and express the vortex transverse size and transverse
energy in terms of $\Lambda_{\overline{MS}}$, see Table ~\ref{resdetails}.
The appearance of the ``11/3'' is actually a powerful test of numerical
precision. Explicit calculations for different transverse sizes confirm the
natural behavior of quantum energy, which follows from dimensions. 
In addition we verify that the quantum energy is proportional to $N_c$, 
as it should be. In our approach, it implies non-trivial sum rules for 
the phase shifts for gluons and ghosts with different color polarizations. \\

The effective transverse energy of a vortex (classical plus quantum)
has a minimum {\em lower} than the perturbative vacuum both in $4d$ and
$3d$. In $4d$ this result is in a qualitative agreement with the well-known 
Savvidy's logarithm, 
$(11N_c/96\pi^2) H^2 \ln(H/\Lambda^2\,{\rm const.})$, 
for the energy of a constant chromomagnetic field $H$, 
following from asymptotic freedom \cite{Sav}. However,
the constant chromomagnetic background has a negative mode \cite{NNO,AO},
so that Savvidy's vacuum is unstable. [Because of this negative mode
the fluctuation determinant is not well-defined, and the constant 
in the argument of the logarithm can not be determined.] In contrast,
the vortex has a fast-varying magnetic field, the fluctuation
determinant has no negative mode and can be computed. Qualitatively, the
energy gain is expected from dimensional analysis and asymptotic freedom (see
section 4) but its value is {\it a priori} unknown. In $3d$ one does not have
the argument of the asymptotic freedom formula, in favor of the energy gain.
Therefore, the fact that we find an energy gain for vortices both in $3d$ and
$4d$ is non-trivial. The center (flux-1) vortices give a larger energy gain
than the non-topological flux-0 ones and from that point of view are
dynamically preferred. This was also the conclusion of the Copenhagen group
\cite{AO} who studied semi-classically a $2d$ hexagonal lattice made of
parallel flux tubes. However, from our calculation we cannot exclude that
fluxes of other magnitude, e.g. flux-2 (which are topologically equivalent to
flux-0 ones) are even more preferred.\\

Despite different shapes of vortex profiles considered (one
exponential, the other power-like) we get close numbers for the energy gain.
It indicates that vortices are `soft' with respect to their deformations in
the transverse plane. A further indication in favor of this interpretation is
the fast rise of the aggregate phase at small momenta, see
Figs.~\ref{aphflux0},\ref{inve3},\ref{invp6}, although
zero modes in a strict sense do not seem to exist, see Appendix B.
That makes the search for the absolute minimum in the 1-loop
approximation rather problematic: it will be very flat anyhow.
Simultaneously, it makes the naive interpretation of thick vortices as
neatly shaped spaghetti even more difficult. \\

Assuming that the vacuum is unstable with respect to the spontaneous
production of individual center vortices, their creation can be, in principle,
stabilized by i) kinetic energy of vortex bending, ii) repulsive
interaction of neighbor vortices and/or of different segments of the
same vortex. Concerning vortex interactions, a $Z(N)$ vortex in a regular
gauge necessarily has a nonzero azimuthal component of the Yang--Mills field,
decaying as $1/\rho$ at large distances from the center in the transverse
plane. Two such vortices will then have unacceptable strong repulsion
unless they are `gauge-combed' in such a way that $A_\phi$ is zero at
large distances.  In such a gauge, however, one necessarily gets a
Dirac sheet (in $3d$) or a Dirac 3-volume (in $4d$) of gauge
singularities. Dirac singularities from two vortices intersect along
a line in $3d$ or along a surface in $4d$. An intersection of two
gauge singularities in non-Abelian theory is, generally speaking, not
a pure gauge but a physical singularity. It is not clear how to avoid
those infinities. In any case, vortex interactions are far from
being an easy topic, but they need to be clarified before the
statistical mechanics of vortices can tell us that the `entangled
spaghetti' vacuum has a preferred free energy.  \\
\begin{appendix}
\section{Conversion into cylindrical coordinates}
\label{app1}
The conversion into cylindrical coordinates is given by the transformations:
\begin{eqnarray}
x = \rho \cos\phi \qquad  y = \rho \sin \phi
\end{eqnarray}
and
\begin{eqnarray}
A_x &=& \cos \phi A_\rho - \sin \phi A_\phi
\nonumber \\
A_y &=& \sin \phi A_\rho + \cos \phi A_\phi
\end{eqnarray}
\begin{eqnarray}
\partial_ x &=& \cos \phi \partial_\rho - \frac{\sin \phi}{\rho}
\partial_\phi
\nonumber \\
\partial_y &=& \sin\phi \partial_\rho +\frac{\cos \phi}{\rho}
\partial_\phi \;.
\end{eqnarray}
One finds:
\begin{eqnarray}
F_{12} &=& \partial_x A_y - \partial_y A_x
\nonumber \\
&=&\left( \cos \phi \partial_\rho - \frac{\sin \phi}{\rho}\partial_\phi
\right)\left(\sin \phi A_\rho + \cos \phi A_\phi\right)
\nonumber \\
&&-
\left(\sin \phi \partial_\rho +\frac{\cos \phi}{\rho}\partial_\phi
\right)
\left(\cos \phi A_\rho - \sin \phi A_\phi\right)
\nonumber \\
&=& \frac{A_\phi}{\rho} +  \frac{\partial A_\phi}{\partial \rho} \;,
\end{eqnarray}
using $A_\rho = 0$, and with $A_\phi = \mu(\rho)/\rho$ one obtains:
\begin{equation}
F_{12} = \mu'(\rho)/\rho \quad
\Rightarrow
\quad (F_{\mu\nu})^2 = 2 [\mu'(\rho)/\rho]^2 \;.
\end{equation}

\section{Excluding possible bound states at $k=0$}

The spectrum of the ghost and gluon operators discussed above is
incomplete if there exist bound states at $k_\perp=0$. Were the vortices
an exact solution of the classical equation of motion, zero modes
would be inevitable. However, our vortices are minima of the
effective action and not of the classical one, therefore there are no
special reasons for zero modes of the small-oscillation operators. \\

To check whether such states exist we substitute $\rho\to t=1/\rho$
in the gluon operator and obtain the reflected differential equation:
\begin{equation}
-t \frac{\partial}{\partial t} t \frac{\partial}{\partial t}
Z(1/t) + \left[
(m-\alpha\mu(1/t))^2 + 2\alpha \beta  \frac{\mu'(1/t)}{t}
\right]Z(1/t) = 0\;.
\end{equation}
$\alpha$ and $\beta$ are two constants depending on the color and space
polarizations of the state under consideration.
Using the notation B(t) = Z(1/t) we take the boundary conditions for a bound
state  $B(0) = 0$. This tells only that B has to be regular
at the origin. Now we can check the asymptotic of the equation using
an explicit flux-1 profile, for example, $\mu(\rho) = \exp(-1/\rho^3)$.
One obtains asymptotically for large $\rho$ and correspondingly for small $t$:
\begin{equation}
\left[-\frac{1}{t}
\frac{\partial}{\partial t} t  \frac{\partial}{\partial t}
 +  \frac{(m-\alpha)^2}{t^2}\right] B_0(t) = 0\;.
\label{asdiffeq}
\end{equation}
For this equation we require power series in $t$. The only non-trivial
regular solution to this is $B_0 (t) = A  t^{|m-\alpha|}$. Therefore, we
use the ansatz:
\begin{equation}
B(t) = A(t) t^{|m-\alpha|}\;.
\end{equation}
One obtains the differential equation for the amplitude:
\begin{equation}
\left[
-\frac{1}{t} \frac{\partial}{\partial t} t \frac{\partial}{\partial t}
- 2\frac{|m-\alpha|}{t}\frac{\partial }{\partial t}
 + \frac{(m-\alpha \mu(1/t))^2-(m-\alpha)^2  + 2\alpha
\beta \mu'(1/t)/t}{t^2}\right]A(t) = 0\;.
\end{equation}
The boundary conditions are now $A(0) = 1, A'(0) = 0$ to make sure
that the asymptotical behavior is correct. Explicit calculation shows
that these states behave as:
\begin{equation}
Z_m^{(\alpha,\beta)}(\rho) \sim \left\{
\begin{array}{ccc}
\rho^{-|m|} & {\rm for\;small}& \rho \\
\rho^{-|m-\alpha|} & {\rm for\;large}& \rho \\
\end{array}
\right. \;{\rm for}\; m>0 \;.
\end{equation}
The behavior at small $\rho$ is clear because of asymptotic arguments
like the one used for small $t$ in eq.~(\ref{asdiffeq}). The explicit
calculation shows  that only irregular solutions exist for $m>0$. In the
case $m=0$ the solution
is at least logarithmically divergent for large $\rho$, as $|\alpha|=1,1/2$.
Therefore, there are no localized solutions with zero $k_\perp$.

\end{appendix}


\begin{thebibliography}{99}
\bibitem{footnote}
The only possible loophole in this analysis, known as `Derick's theorem',
is if the field strength of the vortex is allowed to decrease weakly (as
$1/\rho$) at large transverse distances. Then the variation of the energy
functional with size could have a full-divergence piece which potentially
could destroy the argument. [This is what happens in the case of $3d$ BPS
monopoles.] Such behavior, however, would lead to the divergent transverse
energy. Thus, the conclusion that there are no vortex-type solutions of 
the YM classical equations holds true. In a compactified space the situation
is different. In ref. \cite{GAM} a vortex-like classical solution has been
obtained numerically in a partially compactified $R^2\times T^2$ space with
the torus circumference much less than the size of the $R^2$ box. In this case
the scale is set by the compactification circumference. The radius of the
vortex has been found to be of the order of the (short) length in the compact
direction \cite{GAM}.
\end{thebibliography}
\end{document}